\documentclass{jetpl}
\twocolumn
\title{Quantum electrodynamics with anisotropic scaling:
Heisenberg-Euler action and Schwinger pair production in the
bilayer graphene}

\rtitle{Quantum electrodynamics with anisotropic scaling: Schwinger pair production in the
bilayer graphene}

\sodtitle{Quantum electrodynamics with anisotropic scaling:
Heisenberg-Euler action and Schwinger pair production in the
bilayer graphene}

\author
{M. I. Katsnelson$^{*}$ \/\thanks{e-mail: m.katsnelson@science.ru.nl,volovik@boojum.hut.fi}
and G.E. Volovik $^{+\#}$
}

\rauthor{
M.I. Katsnelson, G.E.Volovik
}

\sodauthor{
Katsnelson, Volovik
}

\address{
$^{*}$ Radboud University Nijmegen, Institute for Molecules and Materials, Heyndaalseweg 135, NL-6525AJ Nijmegen, The Netherlands\\
$^{+}$ Low Temperature Laboratory, Aalto University, School of Science and
Technology, P.O. Box 15100, FI-00076 AALTO, Finland
\\
$^{\#}$ Landau Institute for Theoretical Physics RAS, Kosygina 2, 119334 Moscow, Russia
}

\abstract{We discuss quantum electrodynamics emerging in the vacua with
anisotropic scaling. Systems with anisotropic scaling were
suggested by Ho\v{r}ava in relation to the quantum theory of
gravity. In such vacua the space and time are not equivalent, and
moreover they obey different scaling laws, called  the anisotropic
scaling. Such anisotropic scaling takes place for fermions in
bilayer graphene, where if one neglects the trigonal warping
effects the massless Dirac fermions have quadratic dispersion.
This results in the anisotropic quantum electrodynamics, in which
electric and magnetic fields obey different scaling laws. Here we
discuss the Heisenberg-Euler action and Schwinger pair production
in such anisotropic QED.  }

\begin{document}

\maketitle

\section{Introduction}

Both superfluid $^3$He-A \cite{Volovik2003} and single-layer
graphene \cite{CastroNeto2009,Vozmediano2010,Katsnelson2012} serve
as examples of emerging relativistic quantum field theory in 3+1
and 2+1 dimensions, respectively.  Both systems contain
``relativistic'' fermions, which are protected by the combined
action of symmetry and topology (see review \cite{Volovik2011}),
while the collective modes and deformations provide the effective
gauge and gravity fields acting on these fermions
\cite{Volovik2003,Vozmediano2010}.

Effective electromagnetic field emerging in superfluid $^3$He-A is
the collective field which comes from the degrees of freedom
corresponding to the shift of the position of the Weyl (conical)
point in momentum space. The corresponding Maxwell action (for
special orientation of the effective electric and magnetic fields,
see details in \cite{Volovik2003}) is
\begin{equation}
S \sim \int d^3x dt \frac{v_F}{24\pi^2}\left[ B^2-\frac{1}{v_F^2} E^2\right] \ln \frac{1}{\left[ B^2-\frac{1}{v_F^2} E^2\right] }\,,
\label{3He-A}
\end{equation}
where $v_F$ is the Fermi velocity. This action describes the
effect of vacuum polarization, and corresponds to the
Heisenberg-Euler action of standard quantum electrodynamics
\cite{HeisenbergEuler1936}. The only
difference from the latter is that the fermions in the vacuum  of
$^3$He-A are massless, and their masslessness is protected by
momentum-space topology of the Weyl point. This results in the
logarithmic term which describes the zero-charge effect -- the
screening of the effective charge. The effective action becomes imaginary
when $E>v_F B$ describing Schwinger pair production in
electric field \cite{Schwinger1951} with the production rate
$\propto E^2$ for the case of massless fermions \cite{Volovik1992}
(Schwinger pair production of massive fermions has been discussed
for the other topological superfluid -- $^3$He-B, see Ref.
\cite{SchopohlVolovik1992}). Similar action, but for the real
electromagnetic field, should take place in the Weyl semimetals
(for semimetals with the topologically protected Weyl  points see
Refs.
\cite{Abrikosov1971,Nielsen1983,Abrikosov1998,XiangangWan2011,Burkov2011,Aji2011}).
Among the other things, the massless fermions give rise to the
chiral anomaly, and the corresponding Adler-Bell-Jackiw equation
for anomaly has been verified in the $^3$He-A experiments
\cite{Bevan1997}.

Graphene gives the opportunity not only to study the 2+1 quantum
electrodynamics, but also extend the theory to different
directions. In particular, to study systems with anisotropic
scaling, which were suggested by Ho\v{r}ava for construction of
the quantum theory of gravity, which does not suffer the
ultraviolet (UV) divergencies (the  UV completion of general
relativity)
\cite{HoravaPRL2009,HoravaPRD2009,Horava2008,Horava2010}. As
distinct from the relativistic massless fermions in a single layer
graphene, which obey the invariance under conventional scaling
${\bf r}\rightarrow b {\bf r}$, $t\rightarrow b t$, fermions in a
bilayer ($N=2$) or rhombohedral $N$-layer ($N>2$) graphene have a
smooth touching of the Dirac point, $E^2 \propto p^{2N}$, see Refs.
\cite{CastroNeto2009,Katsnelson2012}. These fermions obey the
anisotropic scaling ${\bf r}\rightarrow b {\bf r}$, $t\rightarrow
b^N t$, precisely which is needed for construction of the
divergence-free quantum gravity.

Here we discuss the effect of this anisotropic scaling on the
effective action for real or artificial (e.g., created by
deformations \cite{Vozmediano2010,Katsnelson2012}) electromagnetic
fields. Due to anisotropic scaling, which distinguishes between
the space and time components, the electric and magnetic fields
experience different scaling laws. In particular, the one-loop
action for the magnetic field is $\propto B^2 p_0^{N-4+D}$, where
$D$ is the space dimension and $p_0$ is the infrared cut-off. This
demonstrates that superfluid $^3$He-A with its $D=3$ and $N=1$ and
the bilayer graphene with its $D=N=2$ both correspond to the
critical dimension $D_c=4-N$ and thus they both give rise to the
logarithmically divergent action for the magnetic field (which
manifests itself as a logarithmic divergence of diamagnetic
susceptibility in undoped bilayer graphene with the trigonal
warping effects being neglected \cite{safran,koshino}). However,
for the electric field the action is $\propto E^2 p_0^{D-2-N}$.
While for the superfluid $^3$He-A this again gives the logarithmic
divergence as it happens in the relativistic systems with massless
fermions, in the multiple-layered graphene with its $D=2$ the
action diverges in the infrared for any $N>1$, giving rise to the
power-law action $\propto E^{(N+2)/(N+1)}$.

\section{Effective action for real and induced electromagnetic field}
 \label{sec:EffectiveAction}

{\it Anisotropic scaling for electromagnetic field}
 \label{sec:HoravaGravity}

The effective Hamiltonian for the bilayer graphene in the simplest
approximation is \cite{CastroNeto2009,Katsnelson2012}
\begin{equation}
{\cal H}= \frac{\sigma^+}{2m}\left((\hat{\bf x}+i\hat{\bf y})\cdot({\bf p}- e{\bf A}) \right)^2
+
 \frac{\sigma^-}{2m}\left((\hat{\bf x}-i\hat{\bf y})\cdot({\bf p}- e{\bf A})\right)^2\,,
\label{FermionHamiltonian2}
\end{equation}
where $\sigma$ are Pauli matrices and $m$ is the mass entering the
quadratic band touching. Experimentally, $m \approx 0.03 m_e$
where $m_e$ is the free-electron mass \cite{mayorov}. Here we
neglect the degrees of freedom related to the tetrad gravity and
concentrate on the degrees of freedom corresponding to the
electromagnetic field (in principle the vector potential ${\bf A}$
may include not only the real electromagnetic field, but also the
collective field which come from the degrees of freedom
corresponding to the shift of the position of the Dirac point in
momentum space, as in $^3$He-A). We shall use the natural units in
which $\hbar=1$; electric charge $e=1$; the vector field ${\bf A}$
has dimension of momentum, $[A]=[L]^{-1}$; electric and magnetic
fields have dimensions $[E]=[LT]^{-1}$ and $[B]=[L]^{-2}$,
respectively.

For standard quantum electrodynamics emerging in the vacuum with massive
electrons, the Lorentz invariance combined with the dimensional
analysis gives the general form $(B^2-E^2)f(x,y)$ for the
Heisenberg-Euler action \cite{HeisenbergEuler1936} in terms of
dimensionless quantities  $x=(B^2-E^2)/M^4$, $y= {\bf B}\cdot {\bf
E}/M^4$. Here $M$ is the rest energy of electron, which violates
the scale invariance in the infrared. Extension of the Born-Infeld electrodynamics to the 
anisotropic scaling has been considered in Ref. \cite{Andreev2010}. 
We consider the electrodynamics with anisotropic scaling, which is induced by 
fermions obeying the Hamiltonian \eqref{FermionHamiltonian2}.

{\it General effective action}
 \label{sec:GenEffectiveAction}

In Eq. \eqref{FermionHamiltonian2} there is only one dimensional
parameter, the mass $m$. Combining the dimensional analysis with
the anisotropic scaling, one obtains that the effective action for
the constant in space and time electromagnetic field, which is
obtained by the integration over the 2+1 fermions with quadratic
dispersion, is the function of the dimensionless combination $\mu$
of electric and magnetic field
 \begin{equation}
S = \int d^2x dt  ~\frac{B^2}{m} g(\mu)~~,~~ \mu=\frac{m^2E^2}{B^3}\,.
\label{EffectiveActionGeneral}
\end{equation}
The change of the regime from electric-like to magnetic-like
behavior of the action occurs at $\mu \approx 1$. The asymptotical
behavior in two limit cases, $g(\mu \rightarrow 0)=a$ and $g(\mu
\rightarrow \infty)= (b+ic)\mu^{2/3}$, gives the effective actions
for the constant in space and time magnetic and electric fields:
 \begin{equation}
S_{B}=a \int d^2x dt  ~\frac{B^2}{m} ~~,~~ S_{E}= \int d^2x dt   (b+ic) E^{4/3}m^{1/3}\,.
\label{BandE}
\end{equation}
The dimensionless parameters $a$ and $b$ describe the vacuum
polarization, and the dimensionless parameter $c$ describes the
instability of the vacuum with respect to the Schwinger pair
production in the electric field, which leads to the imaginary
part of the action.

This should be contrasted with the single layer graphene, where the corresponding
effective action for the 2+1 relativistic quantum electrodynamics is
\begin{equation}
S \sim \int d^2x dt ~v_F\left[ B^2-\frac{1}{v_F^2} E^2\right]^{3/4}\,,
\label{SingleLayer}
\end{equation}
critical field is $E_c(B) = v_F B$; and the Schwinger pair
production is $\propto E^{3/2}$ at $B=0$ (see Refs.
\cite{AndersenHaugset1995,allor,Vildanov2009}). Returning to the
conventional units, one should substitute $B\rightarrow eB/c$ and
$E\rightarrow eE$.

{\it Action for magnetic field}
 \label{sec:Bfield}

First, let us consider the case $E=0$. The action is
\begin{equation}
S=a\int d^2xdt\frac{B^2}m
\end{equation}
where $a$ is undefined yet numerical coefficient. The simplest way
to find $a$ is the following. First, let us consider the case of
uniform magnetic field and the sample of unit area, than, $\int
d^2x\rightarrow 1$. Second, let us go to the imaginary time, $\exp
\left( iS\right) \rightarrow \exp \left( -\beta F\right) $ where
$F$ is the free energy, $\beta =1/T$ is the inverse temperature.
The change of the free energy in magnetic field is $F=-\chi B^2/2$
where $\chi $ is the magnetic susceptibility (per unit area).
Thus, we have
\begin{equation}
a=\frac{m\chi }2
\end{equation}

The susceptibility for the case of bilayer within the model of
purely parabolic spectrum has been calculated in Refs.
\cite{safran,koshino}. As we know, the $D=2$ vacuum of fermions
with quadratic touching, $N=2$, corresponds to the critical
dimension for the magnetic field action: it is logarithmically
divergent for the case of zero doping. This logarithm should be
cut at the energy of the reconstruction of the spectrum due to the
farther hopping effects (``trigonal warping'') and/or
interelectron interaction (for the most recent discussion, see
Ref. \cite{mayorov}).  The answer is
\begin{equation}
a=\frac{g_sg_v}{32\pi }\Lambda \,,
\end{equation}
where $g_s=g_v=2$ are spin and valley degeneracies.
For small fields, the logarithmically running coupling is
 $\Lambda =\ln \left( \gamma _1^2/\Delta^2 \right) $, where
the ultraviolet cut-off is provided by the interlayer hopping
$\gamma _1$, while the infrared cut-off is provided by the
trigonal warping whose typical energy is  $\Delta \approx
0.01\gamma _1$. For larger fields, $B>\Delta^2$,  the infrared
cut-off is provided by the field itself,
 $\Lambda =\ln \left( \gamma _1^2/B \right)$, which is similar to what occurs in effective electrodynamics
 in $^3$He-A with massless fermions, see Eq.\eqref{3He-A}.

{\it Schwinger pair production}
 \label{sec:Schwinger}

Schwinger pair production in bilayer graphene in zero magnetic
field has been considered in Ref. \cite{Vildanov2009}. Let us
consider the case of the crossed fields ${\bf B}=B\hat{\bf z}$,
${\bf E}=E\hat{\bf x}$ using the semiclassical approximation.  We
will use the gauge $A_x=0,A_y=Bx$. Thus, the imaginary part of the
momentum along $x$ direction is determined by the equation (cf.
Ref. \cite{Vildanov2009}):
\begin{equation}
\kappa ^2\left( x\right) =\left( k+Bx\right) ^2-2mE\left| x\right|
\end{equation}
where $k$ is the (conserving) momentum in $y$-direction. The
classically forbidden regions relevant for the tunneling is determined by the condition $%
\kappa ^2\left( x\right) >0$ and the tunneling exponent is
determined by the imaginary part of the
action
\begin{equation}
S_E\left( k\right) =2\int\limits_{x_L}^{x_R}dx\kappa \left(
x\right)
\end{equation}
where $x_L,x_R^{}$ are the left and right turning points.

One has to introduce the parameter
\begin{equation}
\mathcal{E}=\frac{mE}B
\end{equation}
and
\begin{equation}
X=k+Bx
\end{equation}
thus,
\begin{equation}
S_E\left( k\right) =\frac 2B\int\limits_{X_L}^{X_R}dX\kappa \left(
X\right)
\end{equation}
where
\begin{equation}
\kappa ^2\left( X\right) =\left\{
\begin{array}{cc}
\left( X-\mathcal{E}\right) ^2+\mathcal{E}\left( 2k-\mathcal{E}\right) , &
X>k \\
\left( X+\mathcal{E}\right) ^2-\mathcal{E}\left( 2k+\mathcal{E}\right) , &
X<k
\end{array}
\right.
\end{equation}
Now we have to study conditions of existence of the left and right
turning points. We can restrict ourselves by the case $k>0$ only,
due to symmetry with respect to the replacement $k\rightarrow
-k,x\rightarrow -x$. Simple analysis show that two turning point
exist only if
\begin{equation}
\mathcal{E}>2k
\end{equation}
The further calculations are straightforward, and the answer is
\begin{equation}
S_E\left( k\right) =\mu f\left( \frac{2k}{\mathcal{E}}%
\right)
\end{equation}
where
\begin{equation}
f\left( x\right) =x-\frac{1+x}2\ln \left( 1+x\right)
-\frac{1-x}2\ln \left( \frac 1{1-x}\right)
\end{equation}
The Taylor expansion for the function $f$ reads
\begin{equation}
f\left( x\right) = \sum_{n=1}^{\infty} \frac{x^{2n+1}}{2n(2n+1)}
\end{equation}

For $\mathcal{E}\rightarrow \infty $ we have $f\left( x\right)
\approx x^3/6 $, which is in agreement with Ref.
\cite{Vildanov2009}, if one fixes the misprint in  Eq.(36) of
\cite{Vildanov2009}, where there is a power 1/3 instead of 3. The
semiclassical approximation is valid if $\mu \gg 1$.

The pair production is obtained after integration of the tunneling exponent over the momentum
$k$:
 \begin{eqnarray}
{\rm Im}~ S  =\frac{g_sg_v}{2\pi^2} E \int_{0}^{\mathcal{E}/2} dk
\exp\left(- S_E\left( k\right)\right)
\nonumber
\\
=\frac{g_sg_v}{4\pi^2} B^2\mu
\int_{0}^{1} dx \exp\left(- \mu f(x)\right)
 \,.
\label{ImaginaryAction}
\end{eqnarray}
Semiclassically, there is \textit{always} some states which
demonstrate tunneling, but only with $\left| k\right|
<\mathcal{E}/2$. In the case of strong magnetic field their contribution shrinks to the point.

In the limit of small magnetic fields one obtains the parameter $c$ in the action \eqref{BandE}:
\begin{eqnarray}
\nonumber
{\rm Im} S= \frac{g_sg_v}{4\pi^2} B^2\mu  \int_{0}^{+\infty} dx
\exp\left(- \frac{\mu}{6} x^3\right)=cE^{4/3}m^{1/3}\,,
\\
c=\frac{g_sg_v}{12\pi^2}6^{1/3}\Gamma(1/3) .
\label{ImaginaryActionE}
\end{eqnarray}
The Schwinger pair production rate proportional to $E^{4/3}$ has
been discussed in Ref. \cite{Vildanov2009}.

If we take into account the next-order corrections in $1/\mu$ the
expression \eqref{ImaginaryActionE} is multiplied by the factor
$1-\frac{3^{5/3} 2^{2/3}}{10 \Gamma(1/3)}\frac{B^2}{(mE)^{4/3}}$.

The change of the regime from magnetic-like to electric-like
happens at $\mu \approx 1$ which means, in CGSE units,
\begin{equation}
\frac{E}{B} \approx \frac{\hbar}{mc}\sqrt{\frac{|e|B}{\hbar c}}
\end{equation}
For the field $B = 1$T this ratio is of the order of $3 \cdot
10^{-3}$, that is, the order of magnitude smaller than for the
case of the single-layer graphene where it is $v_F/c \approx
1/300$.

{\it Effective action for higher order touching}
 \label{sec:multilayered}

All this can be extended to the order of $N$ band touching, which
presumably may be achieved by rhombohedral stacking of $N$
graphene-like layers \cite{CastroNeto2009,Katsnelson2012}. The
effective Hamiltonian for fermions in this case is
\begin{equation}
{\cal H}= \frac{\sigma^+}{2m}\left((\hat{\bf x}+i\hat{\bf y})\cdot({\bf p}- e{\bf A}) \right)^N
+
 \frac{\sigma^-}{2m}\left((\hat{\bf x}-i\hat{\bf y})\cdot({\bf p}- e{\bf A})\right)^N.
\label{FermionHamiltonianN}
\end{equation}
If we are interested only in the infrared behavior of the action
in the regime when the electric field dominates, then the induced
electromagnetic action has the following general structure:
\begin{equation}
S(N) = \int d^2x dt~  \frac{B^{(2+N)/2}}{m} g(\mu)~~,~~ \mu=\frac{m^2E^2}{B^{N+1}}\,.
\label{EMactionGeneral}
\end{equation}
At large $\mu$ one has $g(\mu) \propto \mu^{(N+2)/2(N+1)}$, and
the Schwinger pair production rate in zero magnetic field is
 \begin{equation}
\dot n \sim  \frac{1}{m}\left(mE\right)^{\frac{N+2}{N+1}} .
\label{SchwingerN}
\end{equation}

\section{Discussion}
 \label{sec:Discussion}

In the condensed matter context (or in related microscopic
theories of the quantum vacuum), the existence of the
topologically protected nodes in spectrum of Weyl media gives rise
to the effective gauge fields ($U(1)$ and $SU(2)$) and gravity.
The same takes place in single layer and bilayer graphene, which
contain Dirac points in their spectrum. The effective $SU(2)$
gauge field comes from the spin degrees of freedom, these are the
collective modes in which the momentum of the Dirac point shifts
differently for spin-up and spin-down species. In addition to the
collective modes related to the shift of the Dirac point, the
graphene has degrees of freedom, which correspond to tetrads in
anisotropic gravity.  In particular, for the bilayer graphene
these degrees of freedom enter the effective Hamiltonian in the
following way
\begin{eqnarray}
{\cal H}= \sigma^+\left(({\bf e}_1+i{\bf e}_2)\cdot({\bf p}- e{\bf A}) \right)^2
\nonumber
\\
+
 \sigma^-\left(({\bf e}_1-i{\bf e}_2)\cdot({\bf p}- e{\bf A})\right)^2\,,
\label{Tetrad}
\end{eqnarray}
where ${\bf e}_1$ and ${\bf e}_2$ play the role of  zweibein fields, which give rise to the energy spectrum
$E^2=\left(g^{ik}p_ip_k\right)^2$ corresponding to the effective 2D metric
$g^{ik}=e_1^ie_1^k + e_2^ie_2^k$; to spin connection and torsion.

As distinct from superfluid $^3$He-A, which is the analog of the
relativistic vacuum with massless Weyl fermions, the bilayer
graphene is the representative of the quantum vacua, which
experience different scaling laws for space and time. While such
vacua were considered by  Ho\v{r}ava in relation to quantum
gravity, here we applied the anisotropic scaling to quantum
electrodynamics emerging in these systems, using as an example the
2D system with massless Dirac fermions with quadratic spectrum.
Such systems have peculiar properties, and we touched here the
Heisenberg-Euler  action and the Schwinger pair production.

\section*{Acknowledgements}

It is our pleasure to thank Frans Klinkhamer for discussion.
This work is supported by the Stichting voor Fundamenteel
Onderzoek der Materie (FOM), which is financially supported by the
Nederlandse Organisatie voor Wetenschappelijk Onderzoek (NWO), and
by the Academy of Finland and its COE program.


\begin{thebibliography}{99}

\bibitem{Volovik2003}
G.E. Volovik,
{\it The Universe in a Helium Droplet},
Clarendon Press,  Oxford (2003).

\bibitem{CastroNeto2009} A.H. Castro Neto, F. Guinea, N.M.R. Peres, K.S. Novoselov,
and A.K. Geim, Rev. Mod. Phys. \textbf{81}, 109--162 (2009).

\bibitem{Vozmediano2010}
M.A.H. Vozmediano, M.I. Katsnelson, and F. Guinea,
Physics Reports {\bf 496}, 109--148 (2010).

\bibitem{Katsnelson2012}
M.I. Katsnelson, {\it Graphene: Carbon in Two Dimensions},
Cambridge Univ. Press, Cambridge (2012).

\bibitem{Volovik2011}
G.E. Volovik,
Topology of quantum vacuum,
draft for Chapter in proceedings of the Como Summer School on analogue gravity,
 arXiv:1111.4627.

 \bibitem{HeisenbergEuler1936}
 W. Heisenberg and H. Euler,
 Z. Phys. {\bf 98}, 714--732 (1936).

\bibitem{Schwinger1951}
 J. Schwinger,
  Phys. Rev. {\bf 82}, 664--679  (1951).

\bibitem{Volovik1992}
G.E. Volovik,
{\it Exotic properties of superfluid $^3$He},
World Scientific, Singapore (1992).

\bibitem{SchopohlVolovik1992}
N. Schopohl and G.E. Volovik,
Ann. Phys. (N. Y.) {\bf 215}, 372--385 (1992).


 \bibitem{Abrikosov1971}
A.A. Abrikosov    and S.D. Beneslavskii,
Sov. Phys. JETP {\bf 32}, 699--708 (1971).

 \bibitem{Nielsen1983}
H.B. Nielsen and M. Ninomiya,
Phys. Lett. B {\bf 130},  389--396  (1983).


 \bibitem{Abrikosov1998}
A.A. Abrikosov,
 Phys. Rev. {\bf B 58}, 2788--2794 (1998).

\bibitem{XiangangWan2011}
X. Wan, A.M. Turner,  A. Vishwanath  and S.Y. Savrasov,
Phys. Rev. B {\bf 83}, 205101 (2011).

\bibitem{Burkov2011}
A.A. Burkov and L. Balents,
Phys. Rev. Lett. {\bf 107}, 127205 (2011).

 \bibitem{Aji2011}
 V. Aji,
arXiv:1108.4426.

\bibitem{Bevan1997}
T.D.C. Bevan, A.J. Manninen, J.B. Cook, J.R. Hook, H.E. Hall, T.
Vachaspati, and G.E. Volovik,
Nature {\bf 386},  689--692 (1997).

\bibitem{HoravaPRL2009}
P. Ho\v{r}ava,
Phys. Rev. Lett. {\bf 102}, 161301 (2009).

\bibitem{HoravaPRD2009}
P. Ho\v{r}ava,
Phys. Rev. D {\bf 79}, 084008 (2009).

\bibitem{Horava2008}
P. Ho\v{r}ava,
JHEP 0903, 020 (2009),

\bibitem{Horava2010}
C. Xu and P. Ho\v{r}ava,
Phys. Rev. D {\bf 81}, 104033 (2010).

\bibitem{AndersenHaugset1995}
J.O. Andersen and T. Haugset,
Phys. Rev. D {\bf 51},3073--3080 (1995).

\bibitem{safran}
S.A. Safran,
Phys. Rev. B \textbf{30}, 421--423 (1984).

\bibitem{koshino}
M. Koshino and T. Ando,
Phys. Rev. B \textbf{76}, 085425 (2007).

\bibitem{mayorov}
A.S. Mayorov, D.C. Elias, M. Mucha-Kruczynski, R.V. Gorbachev, T.
Tudorovskiy, A. Zhukov, S.V. Morozov, M.I. Katsnelson, V.I.
Fal'ko, A.K. Geim, and K.S. Novoselov, Science \textbf{333},
860--863 (2011).


\bibitem{Andreev2010}
O. Andreev,
Int. J. Mod. Phys. A {\bf 25}, 2087--2101 (2010);
arXiv:0910.1613.

\bibitem{allor} D. Allor, T.D. Cohen, and D.A. McGady,
Phys. Rev. D {\bf 78}, 096009 (2008).

\bibitem{Vildanov2009}
N.M. Vildanov,
J. Phys.: Condens. Matter {\bf 21}, 445802 (2009).

\end{thebibliography}
\end{document}